\documentstyle[onecolumn,epsf]{mn}

\newcommand{\sI}{{\cal I}}

\title[An alpha theory of warped accretion discs]
  {An alpha theory of time-dependent warped accretion discs}
\author[G. I. Ogilvie]{G. I. Ogilvie$^{1,2}$\\
  $^1$Max-Planck-Institut f\"ur Astrophysik,
  Karl-Schwarzschild-Stra\ss e 1, Postfach 1523,
  D-85740 Garching bei M\"unchen, Germany\\
  $^2$Institute of Astronomy, University of Cambridge, Madingley Road,
  Cambridge CB3 0HA}

\begin{document}

\maketitle

\label{firstpage}

\begin{abstract}
  The non-linear fluid dynamics of a warped accretion disc was
  investigated in an earlier paper by developing a theory of fully
  non-linear bending waves in a thin, viscous disc.  That analysis is
  here extended to take proper account of thermal and radiative
  effects by solving an energy equation that includes viscous
  dissipation and radiative transport.  The problem is reduced to
  simple one-dimensional evolutionary equations for mass and angular
  momentum, expressed in physical units and suitable for direct
  application.  This result constitutes a logical generalization of
  the alpha theory of Shakura \& Sunyaev to the case of a
  time-dependent warped accretion disc.  The local thermal-viscous
  stability of such a disc is also investigated.
\end{abstract}


\section{Introduction}

There are many good reasons for investigating the dynamics of thin
accretion discs that are warped or twisted, in the sense that fluid
elements closely follow circular Keplerian orbits, but the orbital
plane varies slowly with radius and possibly with time.  The existence
of such discs is supported by a growing body of observational evidence
and theoretical reasoning.

Direct observational evidence for warped profiles has been obtained in
exceptional cases\footnote{The more familiar examples of warped spiral
  galaxies are not considered here, because they are likely to be
  controlled by different physical effects.}, including the
circumnuclear disc in the galaxy NGC~4258, revealed through
very-long-baseline interferometric observations of water masers
(Miyoshi et~al. 1995), and the dusty circumstellar disc of $\beta$~Pic
(e.g.~Heap et~al. 2000).  Indirect observational evidence comes from
several low-mass X-ray binaries in which a long periodicity or other
modulation is attributed to a precessing tilted disc (see Wijers \&
Pringle 1999 and references therein).  The best example is the
eclipsing system Her~X-1 (Tananbaum et~al. 1972; Katz 1973; Roberts
1974), but the precessing jets of SS~433 (e.g.~Margon 1984) also
presumably originate in a tilted disc.  Further indirect evidence
comes from pre-main-sequence binaries or unresolved young stellar
objects from which two jets emanate in different directions (see
Terquem et~al. 1999 and references therein).  If the jets are emitted
perpendicular to the discs that produce them, at least one of the
discs must be misaligned with the binary orbital plane.  Such a
misalignment appears to have been captured directly in images of HK
Tau (Stapelfeldt et~al.  1998; Koresko 1998).  In each of these cases,
the disc cannot simply be rigidly tilted, but must become warped under
the influence of the tidal potential.

Even without observational motivation, there are theoretical grounds
for considering warped discs.  In any situation in which, for
historical reasons, the disc is misaligned with the spin axis of a
central black hole (Bardeen \& Petterson 1975), the axis of a central
magnetosphere, the orbit of a companion object (Papaloizou \& Terquem
1995), or the angular momentum vector of matter being supplied at a
large distance, the disc experiences a external torque that attempts
to change its orientation.  The response of the disc is basically
gyroscopic, but, unlike a rigid body, it communicates internally
through bending waves and `viscous' torques.  In these situations the
disc is likely to become warped, at least temporarily.  Even in a
system that is initially coplanar, a warp may develop spontaneously
through instabilities, which may be due to resonant tidal interactions
(Lubow 1992; Lubow \& Ogilvie 2000), wind torques (Schandl \& Meyer
1994), radiation torques (Pringle 1996) or magnetic torques (Lai
1999).

Historically, attempts to describe the dynamics of warped accretion
discs have involved deriving linear equations for slowly varying $m=1$
bending disturbances of an initially flat disc.  It is usually assumed
that the agent responsible for angular momentum transport in accretion
discs (probably magnetohydrodynamic turbulence) may be treated as an
isotropic effective viscosity, described by the dimensionless
parameter $\alpha$.  For Keplerian discs an important qualitative
distinction exists between significantly viscous discs ($\alpha\ga
H/r$, where $H/r$ is the angular semi-thickness of the disc), in which
warps evolve diffusively (Papaloizou \& Pringle 1983), and almost
inviscid discs ($\alpha\la H/r$), in which warps propagate as
approximately non-dispersive waves (Papaloizou \& Lin 1995).

However, a non-linear theory is required in many cases where the disc
is warped sufficiently to have observational consequences.  Different
approaches have been taken to the non-linear problem.  Direct
numerical simulations have been carried out using smoothed particle
hydrodynamics (SPH; e.g.~Larwood et~al. 1996; Nelson \& Papaloizou
1999).  This method makes no attempt to resolve the
magnetohydrodynamic turbulence but involves an artificial viscosity
over which some control can be exerted.  It is most suitable for
relatively thick discs of limited radial extent, and these
investigations have focused mainly on the case $\alpha\la H/r$.
Usually good agreement is found with linear theory, but new,
non-linear effects have also been found.  The computational expense
makes it difficult to follow the evolution of the disc on the long,
viscous time-scale.

Meanwhile, the semi-analytical method has been developed into a
non-linear theory.  Pringle (1992) adopted plausible evolutionary
equations for a thin, viscous disc with a warp of arbitrary amplitude.
The equations are simple conservation laws for mass and (vector)
angular momentum in one (radial) dimension.  The advantages of this
approach are that the equations have a clear physical interpretation
in terms of torques acting on rings, and that the numerical
implementation is much less demanding than for a direct numerical
simulation, so that the evolution of the disc on the viscous
time-scale is easily described.

In an effort to provide a more complete justification of the latter
approach, I have developed a theory of fully non-linear bending waves
in a thin, viscous disc, starting from the Navier-Stokes equation in
three dimensions (Ogilvie 1999, hereafter Paper~I).  It was shown
that, in cases where the resonant wave propagation found in inviscid
Keplerian discs (Papaloizou \& Lin 1995) may be neglected, the problem
can indeed be reduced to one-dimensional evolutionary equations as
suggested by Pringle (1992).  Moreover, the dimensionless coefficients
in the equations, and their variation with the basic parameters of the
disc and with the amplitude of the warp, were derived, and a proper
connection was made with the existing linear theory.  Further
discussion of the meaning and validity of this approach is given in
Section~5.2 below.

It was recognized that the treatment of thermal physics in Paper~I was
oversimplified.  In order to make the non-linear problem tractable, a
polytropic relation was assumed to hold locally in radius.  In
reality, the thickness of the disc and the stresses between
neighbouring rings depend on a balance between the local rate of
dissipation of energy and the radiation from the surfaces of the disc.
In a warped disc, as a result of induced motions, the viscous
dissipation is enhanced where the warp is strongest, and this affects
the non-linear dynamics of the warp.  The aim of the present paper is
to give a proper treatment of thermal physics by solving an energy
equation including viscous dissipation and radiative transport.
Remarkably, the problem can still be reduced by separation of
variables to a set of dimensionless, non-linear ordinary differential
equations (ODEs) which yield the required coefficients.

The remainder of this paper is organized as follows.  I present the
extended set of equations in Section~2 and solve for the instantaneous
azimuthal and vertical structure of the warped disc and the induced
motions.  I then deduce the complete set of evolutionary equations for
the warped disc.  In Section~3 the evolutionary equations are tested
for stability with respect to short-wavelength perturbations.  The
outcome of numerical evaluation of the dimensionless torque
coefficients and the local stability of the disc is presented in
Section~4.  The results are collected in Section~5 in a concise form
suitable for direct application, and the meaning and validity of the
theory are discussed.  In the Appendix, the possible effects of
external irradiation are assessed.

The detailed application of this theory to warped discs in X-ray
binaries will be given in a forthcoming publication (Ogilvie \& Dubus
2000).

\section{The extended equations and their solution}

\subsection{The energy equation and radiative transport}

The basic equations and notation are taken to be the same as in
Paper~I (equations I.37--I.39), except that the adiabatic condition
(equation I.38) is replaced by an energy equation that includes both
viscous dissipation and radiative transport of energy within the
Rosseland approximation.  The governing equations are therefore the
equation of mass conservation,
\begin{equation}
  {{{\rm D}\rho}\over{{\rm D}t}}=-\rho\nabla\!\cdot\!{\bmath u},
\end{equation}
the equation of motion,
\begin{equation}
  \rho{{{\rm D}{\bmath u}}\over{{\rm D}t}}=-\nabla p-\rho\nabla\Phi+
  \nabla\!\cdot\!
  \left[\mu\nabla{\bmath u}+\mu(\nabla{\bmath u})^{\rm T}\right]+
  \nabla\left[(\mu_{\rm b}-{\textstyle{{2}\over{3}}}\mu)
  \nabla\!\cdot\!{\bmath u}\right],
\end{equation}
and the energy equation,
\begin{equation}
  \left({{1}\over{\Gamma-1}}\right){{{\rm D}p}\over{{\rm D}t}}=
  -\left({{\Gamma}\over{\Gamma-1}}\right)p\nabla\!\cdot\!{\bmath u}+
  (\nabla{\bmath u})\!:\!
  \left[\mu\nabla{\bmath u}+\mu(\nabla{\bmath u})^{\rm T}\right]+
  (\mu_{\rm b}-{\textstyle{{2}\over{3}}}\mu)
  (\nabla\!\cdot\!{\bmath u})^2-\nabla\!\cdot\!{\bmath F}.
\label{newenergy}
\end{equation}
Here 
\begin{equation}
  {\bmath F}=-{{16\sigma T^3}\over{3\kappa_{\rm R}\rho}}\nabla T
\end{equation}
is the radiative energy flux, where $\sigma$ is the Stefan-Boltzmann
constant, $T$ the temperature and $\kappa_{\rm R}$ the Rosseland mean
opacity.  The disc is assumed to be optically thick.  The adiabatic
exponent $\Gamma(r)$ is assumed to be a prescribed function of radius,
but a polytropic relation is no longer adopted.  The equation of state
is that of an ideal gas,
\begin{equation}
  p={{k\rho T}\over{\mu_{\rm m}m_{\rm H}}},
\end{equation}
where $k$ is the Boltzmann constant, $\mu_{\rm m}$ the mean molecular
mass (assumed constant) and $m_{\rm H}$ the mass of the hydrogen atom.
As in Paper~I, the dynamic shear and bulk viscosities are assumed to
be locally proportional to the pressure, so that
\begin{eqnarray}
  \mu&=&\alpha p/\Omega,\\
  \mu_{\rm b}&=&\alpha_{\rm b} p/\Omega,
\end{eqnarray}
where the dimensionless coefficients $\alpha(r)$ and $\alpha_{\rm
  b}(r)$ are prescribed functions of radius.  Finally, the opacity law
is assumed to be of the general power-law form
\begin{equation}
  \kappa_{\rm R}=C_\kappa\rho^xT^y,
\end{equation}
where $C_\kappa$ is a constant.  This includes the important cases of
Thomson scattering opacity ($x=0$, $y=0$) and Kramers opacity ($x=1$,
$y=-7/2$).  The appropriate values of $C_\kappa$ are discussed in
Section~5.1 below.

\subsection{Set~A equations}

In Paper~I the equations of fluid dynamics were derived in warped
spherical polar coordinates, which follow the principal warping motion
of the disc, and were then reduced by means of asymptotic expansions
for a thin disc.  In physical terms, this separates the `fast' orbital
motion from the `intermediate' oscillatory velocities induced by the
warp and from the `slow' velocities associated with accretion.  The
equations to be considered form two sets.

Set~A, which determines the `intermediate' velocities, is a set of
coupled non-linear partial differential equations (PDEs) in two
dimensions, azimuthal and vertical.  The physical content of Set~A is
as follows.  In a flat disc, the pressure and viscous stress are
stratified in the vertical direction.  When the disc is warped, this
produces horizontal forces of odd symmetry about the mid-plane.  These
forces drive shearing horizontal epicyclic motions in the disc.  The
horizontal motions couple non-linearly with the warp itself to produce
compressive vertical motions, causing the thickness of the disc to
vary in azimuth.  All the induced motions are subject to viscous
damping and contribute to the energy dissipation rate.  The problem
defined by Set~A is effectively instantaneous and local to a single
annulus; derivatives with respect to time and radius do not appear
because of the separation of scales in a thin disc.  The solution of
Set~A must be obtained in full.

Set~B, which determines the `slow' velocities, is a set of coupled
linear PDEs with coefficients depending on the solution of Set~A.
Set~B does not have to be solved, however, because the necessary
information can be extracted through certain integrations.  This
procedure amounts to obtaining the solvability conditions for the set
of linear equations, and these conditions are precisely the required
equations determining the evolution of the warp.  Physically, they
involve a calculation of the instantaneous internal torque associated
with the `intermediate' velocities and the deformed structure of the
disc, known from the solution of Set~A.

By replacing the adiabatic condition with the energy equation
(\ref{newenergy}), equations (I.69) in Set~A and (I.74) in Set~B are
modified.  However, the manipulations of Set~B in Paper~I did not
involve equation (I.74) and are therefore unaffected.  It is
sufficient to consider the modifications to Set~A.

Recall that ($r,\theta,\phi$) denote warped coordinates, which are
based on spherical polar coordinates, but the equatorial plane
$\theta=\pi/2$ is deformed to correspond to the warped mid-plane of
the disc.  For a thin disc, the scaled dimensionless vertical
coordinate $\zeta$ is used, where $\theta=\pi/2-\epsilon\zeta$ and
$\epsilon\ll1$.  Also $(v_r,v_\theta,v_\phi)$ denote the relative
velocity components, which are proportional to the rates of change of
the three coordinates following a fluid element.  The shape of the
warp is described through the unit tilt vector ${\bmath\ell}(r,t)$,
which has Euler angles $\beta(r,t)$ and $\gamma(r,t)$.  The orbital
angular velocity is $\Omega(r)$ and the associated epicyclic frequency
is $\kappa(r)$.  A prime denotes differentiation with respect to $r$.

In this paper a simplified notation is adopted in which
$\{\rho_0,p_0,\mu_0,\mu_{{\rm b}0},v_{r1},v_{\theta1},v_{\phi1}\}$ are
written as $\{\rho,p,\mu,\mu_{\rm b},v_r,v_\theta,v_\phi\}$.  No
confusion can result because Set~B will not be considered explicitly.
Similarly, the new quantities $\{T,F,\kappa_{\rm R}\}$ represent their
leading-order terms.  Here $F=-F_\theta$ represents the vertical
radiative flux which is the only component present at leading order.

The extended equations of Set~A may be written as follows.  The
equation of mass conservation:
\begin{equation}
  \left(\Omega{{\partial}\over{\partial\phi}}-
  {{v_\theta}\over{r}}{{\partial}\over{\partial\zeta}}\right)\rho=
  {{\rho}\over{r}}{{\partial v_\theta}\over{\partial\zeta}}.
  \label{rho}
\end{equation}
The $r$-component of the equation of motion:
\begin{eqnarray}
  \lefteqn{\rho\left(\Omega{{\partial}\over{\partial\phi}}-
  {{v_\theta}\over{r}}{{\partial}\over{\partial\zeta}}\right)v_r-
  2\rho\Omega(v_\phi+rv_r\gamma^\prime\cos\beta)=
  -(\beta^\prime\cos\phi+\gamma^\prime\sin\beta\sin\phi)
  {{\partial}\over{\partial\zeta}}
  \left[p+\left(\mu_{\rm b}+{{1}\over{3}}\mu\right){{1}\over{r}}
  {{\partial v_\theta}\over{\partial\zeta}}\right]}\nonumber\\
  &&+\left[{{1}\over{r^2}}+
  (\beta^\prime\cos\phi+\gamma^\prime\sin\beta\sin\phi)^2\right]
  {{\partial}\over{\partial\zeta}}
  \left(\mu{{\partial v_r}\over{\partial\zeta}}\right)+
  \Omega(\beta^\prime\sin\phi-\gamma^\prime\sin\beta\cos\phi)
  {{\partial\mu}\over{\partial\zeta}}.
  \label{vr}
\end{eqnarray}
The $\theta$-component:
\begin{eqnarray}
  \lefteqn{\rho\left(\Omega{{\partial}\over{\partial\phi}}-
  {{v_\theta}\over{r}}{{\partial}\over{\partial\zeta}}\right)
  \left[v_\theta+rv_r(\beta^\prime\cos\phi+
  \gamma^\prime\sin\beta\sin\phi)\right]-\rho\Omega\left[r\Omega\zeta+
  rv_r(\beta^\prime\sin\phi-\gamma^\prime\sin\beta\cos\phi)\right]}
  &\nonumber\\
  &&={{1}\over{r}}{{\partial}\over{\partial\zeta}}
  \left[p+\left(\mu_{\rm b}+{{1}\over{3}}\mu\right){{1}\over{r}}
  {{\partial v_\theta}\over{\partial\zeta}}\right]
  +\left[{{1}\over{r^2}}+
  (\beta^\prime\cos\phi+\gamma^\prime\sin\beta\sin\phi)^2\right]
  {{\partial}\over{\partial\zeta}}
  \left\{\mu{{\partial}\over{\partial\zeta}}
  \left[v_\theta+rv_r(\beta^\prime\cos\phi+
  \gamma^\prime\sin\beta\sin\phi)\right]\right\}\nonumber\\
  &&\qquad-r\Omega(\beta^\prime\cos\phi+\gamma^\prime\sin\beta\sin\phi)
  (\beta^\prime\sin\phi-\gamma^\prime\sin\beta\cos\phi)
  {{\partial\mu}\over{\partial\zeta}}.
  \label{vtheta}
\end{eqnarray}
The $\phi$-component:
\begin{eqnarray}
  \lefteqn{\rho\left(\Omega{{\partial}\over{\partial\phi}}-
  {{v_\theta}\over{r}}{{\partial}\over{\partial\zeta}}\right)
  (v_\phi+rv_r\gamma^\prime\cos\beta)+
  {{\rho\kappa^2}\over{2\Omega}}v_r=
  \left[{{1}\over{r^2}}+
  (\beta^\prime\cos\phi+\gamma^\prime\sin\beta\sin\phi)^2\right]
  {{\partial}\over{\partial\zeta}}
  \left[\mu{{\partial}\over{\partial\zeta}}
  (v_\phi+rv_r\gamma^\prime\cos\beta)\right]}\nonumber\\
  &&+r\Omega^\prime(\beta^\prime\cos\phi+\gamma^\prime\sin\beta\sin\phi)
  {{\partial\mu}\over{\partial\zeta}}.
  \label{vphi}
\end{eqnarray}
The energy equation:
\begin{eqnarray}
  \lefteqn{\left({{1}\over{\Gamma-1}}\right)
  \left(\Omega{{\partial}\over{\partial\phi}}-
  {{v_\theta}\over{r}}{{\partial}\over{\partial\zeta}}\right)p=
  \left({{\Gamma}\over{\Gamma-1}}\right){{p}\over{r}}
  {{\partial v_\theta}\over{\partial\zeta}}+
  2\mu\left[(\beta^\prime\cos\phi+\gamma^\prime\sin\beta\sin\phi)
  {{\partial v_r}\over{\partial\zeta}}\right]^2}&\nonumber\\
  &&+2\mu\left\{{{1}\over{r}}{{\partial}\over{\partial\zeta}}
  \left[v_\theta+rv_r(\beta^\prime\cos\phi+
  \gamma^\prime\sin\beta\sin\phi)\right]\right\}^2\nonumber\\
  &&+\mu\left\{(\beta^\prime\cos\phi+\gamma^\prime\sin\beta\sin\phi)
  {{\partial}\over{\partial\zeta}}
  \left[v_\theta+rv_r(\beta^\prime\cos\phi+
  \gamma^\prime\sin\beta\sin\phi)\right]
  -(\beta^\prime\sin\phi-\gamma^\prime\sin\beta\cos\phi)r\Omega-
  {{1}\over{r}}{{\partial v_r}\over{\partial\zeta}}\right\}^2\nonumber\\
  &&+\mu\left[r\Omega^\prime+
  (\beta^\prime\cos\phi+\gamma^\prime\sin\beta\sin\phi)
  {{\partial}\over{\partial\zeta}}
  (v_\phi+rv_r\gamma^\prime\cos\beta)\right]^2
  +\mu\left[{{1}\over{r}}{{\partial}\over{\partial\zeta}}
  (v_\phi+rv_r\gamma^\prime\cos\beta)\right]^2\nonumber\\
  &&+\left(\mu_{\rm b}-{{2}\over{3}}\mu\right)
  \left({{1}\over{r}}{{\partial v_\theta}\over{\partial\zeta}}\right)^2-
  {{1}\over{r}}{{\partial F}\over{\partial\zeta}}.
  \label{energy}
\end{eqnarray}
The radiative flux:
\begin{equation}
  F=-{{16\sigma T^3}\over{3\kappa_{\rm R}\rho}}
  {{1}\over{r}}{{\partial T}\over{\partial\zeta}}.
\end{equation}

Since only derivatives of the dependent variables with respect to
$\phi$ and $\zeta$ appear in this set, their parametric dependence on
$r$ and $t$ is neglected in the notation adopted in the following
analysis.

Recall that the dimensionless complex amplitude of the warp is
\begin{equation}
  \psi=|\psi|{\rm e}^{{\rm i}\chi}=
  r(\beta^\prime+{\rm i}\gamma^\prime\sin\beta).
\end{equation}
The two combinations appearing in Set~A are
\begin{eqnarray}
  r(\beta^\prime\cos\phi+\gamma^\prime\sin\beta\sin\phi)&=&
  |\psi|\cos(\phi-\chi),\\
  r(\beta^\prime\sin\phi-\gamma^\prime\sin\beta\cos\phi)&=&
  |\psi|\sin(\phi-\chi).
\end{eqnarray}

It will be more useful to introduce, instead of the dimensionless
epicyclic frequency $\tilde\kappa$, the quantity
\begin{equation}
  q=-{{{\rm d}\ln\Omega}\over{{\rm d}\ln r}}.
\end{equation}
These are related by $\tilde\kappa^2=4-2q$.

\subsection{Vertical structure of the disc}

It will be instructive to consider first the case of a flat disc.
Then the above equations reduce to the ODEs
\begin{eqnarray}
  {{{\rm d}p}\over{{\rm d}\zeta}}&=&-\rho r^2\Omega^2\zeta,
  \label{flat1}\\
  {{{\rm d}F}\over{{\rm d}\zeta}}&=&q^2\alpha pr\Omega,\\
  {{{\rm d}T}\over{{\rm d}\zeta}}&=&
  -{{3\kappa_{\rm R}\rho r F}\over{16\sigma T^3}},
  \label{flat3}
\end{eqnarray}
subject to the given equation of state, the opacity law, and the
boundary conditions
\begin{equation}
  F=0\qquad\hbox{at $\zeta=0$}
\end{equation}
and
\begin{equation}
  \rho=T=0\qquad\hbox{at $\zeta=\zeta_{\rm s}$},
\end{equation}
where the surface coordinate $\zeta_{\rm s}$ is to be determined.
These `zero boundary conditions' assume that the disc is highly
optically thick and neglect the atmosphere of the disc.  The effects
of finite optical thickness and external irradiation are estimated in
the Appendix.

Recast this problem in a dimensionless form by means of the
transformations
\begin{eqnarray}
  r\zeta&=&\zeta_*\,U_H,\\
  \rho(\zeta)&=&\rho_*(\zeta_*)\,U_\rho,\\
  p(\zeta)&=&p_*(\zeta_*)\,U_p,\\
  T(\zeta)&=&T_*(\zeta_*)\,U_T,\\
  F(\zeta)&=&F_*(\zeta_*)\,U_F,
\end{eqnarray}
where the starred variables are dimensionless and the $U$ quantities
are natural physical units defined by
\begin{equation}
  U_H=(q^2\alpha)^{1/(6+x-2y)}\Sigma^{(2+x)/(6+x-2y)}
  \Omega^{-(5-2y)/(6+x-2y)}
  \left({{\mu_{\rm m}m_{\rm H}}\over{k}}\right)^{-(4-y)/(6+x-2y)}
  \left({{16\sigma}\over{3C_\kappa}}\right)^{-1/(6+x-2y)},
\end{equation}
together with
\begin{eqnarray}
  U_\rho&=&\Sigma U_H^{-1},\\
  U_p&=&\Omega^2U_H^2U_\rho,\\
  U_T&=&\Omega^2\left({{\mu_{\rm m}m_{\rm H}}\over{k}}\right)U_H^2,\\
  U_F&=&q^2\alpha\Omega U_HU_p,
\end{eqnarray}
where
\begin{equation}
  \Sigma=\int_{-\zeta_{\rm s}}^{\zeta_{\rm s}}\rho\,r\,{\rm d}\zeta
\end{equation}
is the surface density.  The dimensionless equations are
\begin{eqnarray}
  {{{\rm d}p_*}\over{{\rm d}\zeta_*}}&=&-\rho_*\zeta_*,\\
  {{{\rm d}F_*}\over{{\rm d}\zeta_*}}&=&p_*,\\
  {{{\rm d}T_*}\over{{\rm d}\zeta_*}}&=&-\rho_*^{1+x}T_*^{-3+y}F_*,\\
  p_*&=&\rho_*T_*,
\end{eqnarray}
subject to the boundary conditions
\begin{equation}
  F_*(0)=\rho_*(\zeta_{{\rm s}*})=T_*(\zeta_{{\rm s}*})=0.
\end{equation}
The definition of $\Sigma$ also requires that
\begin{equation}
  \int_{-\zeta_{{\rm s}*}}^{\zeta_{{\rm s}*}}\rho_*(\zeta_*)
  \,{\rm d}\zeta_*=1.
\end{equation}
For physical values of $x$ and $y$ it may be assumed that a unique
solution exists, and can be obtained numerically.  Let
\begin{equation}
  \sI_*=\int_{-1}^1\rho_*(\zeta_*)\,\zeta_*^2\,{\rm d}\zeta_*
\end{equation}
be the corresponding dimensionless second vertical moment of the
density.  Then the dimensional quantity is
\begin{equation}
  \sI=\int_{-\zeta_{\rm s}}^{\zeta_{\rm s}}\rho r^2\zeta^2
  \,r\,{\rm d}\zeta=\sI_*\,\Sigma U_H^2.
\end{equation}
The numerical solution for Thomson opacity yields $\zeta_{{\rm
    s}*}\approx2.383$ and $\sI_*\approx0.6777$, while that for Kramers
opacity gives $\zeta_{{\rm s}*}\approx2.543$ and $\sI_*\approx0.7094$.

This solution can now be generalized to describe the azimuthal and
vertical structure of a warped disc.

\subsection{Solution of Set~A equations}

In a warped disc, although many additional effects appear, the
equations of Set~A can be satisfied by functions having basically the
same vertical structure as in the flat disc.  However, the
coefficients are different and the thickness of the disc also depends
on $\phi$.  As an intermediate stage to solving Set~A, it is proposed
that, instead of equations (\ref{flat1})--(\ref{flat3}), the relations
\begin{eqnarray}
  {{\partial p}\over{\partial\zeta}}&=&
  -f_2(\phi-\chi)\rho r^2\Omega^2\zeta,\\
  {{\partial F}\over{\partial\zeta}}&=&
  f_1(\phi-\chi)q^2\alpha pr\Omega,\\
  {{\partial T}\over{\partial\zeta}}&=&
  -{{3\kappa_{\rm R}\rho r F}\over{16\sigma T^3}},
\end{eqnarray}
hold, subject to the same boundary conditions as above, where $f_1$
and $f_2$ are dimensionless functions to be determined.  Physically,
$f_1$ reflects the fact that the dissipation rate is enhanced, in a
non-axisymmetric way, above its Keplerian value because of the
additional shear associated with the induced motions.  Similarly,
$f_2$ represents changes to the effective vertical acceleration in the
disc, caused by inertial forces associated with the induced motions.

These relations can be solved as above to give
\begin{eqnarray}
  r\zeta&=&\zeta_*\,
  \left[f_1(\phi-\chi)\right]^{1/(6+x-2y)}
  \left[f_2(\phi-\chi)\right]^{-(3-y)/(6+x-2y)}U_H,\\
  \rho(\zeta)&=&\rho_*(\zeta_*)\,
  \left[f_1(\phi-\chi)\right]^{-1/(6+x-2y)}
  \left[f_2(\phi-\chi)\right]^{(3-y)/(6+x-2y)}U_\rho,\\
  p(\zeta)&=&p_*(\zeta_*)\,
  \left[f_1(\phi-\chi)\right]^{1/(6+x-2y)}
  \left[f_2(\phi-\chi)\right]^{(3+x-y)/(6+x-2y)}U_p,\\
  T(\zeta)&=&T_*(\zeta_*)\,
  \left[f_1(\phi-\chi)\right]^{2/(6+x-2y)}
  \left[f_2(\phi-\chi)\right]^{x/(6+x-2y)}U_T,\\
  F(\zeta)&=&F_*(\zeta_*)\,
  \left[f_1(\phi-\chi)\right]^{(8+x-2y)/(6+x-2y)}
  \left[f_2(\phi-\chi)\right]^{x/(6+x-2y)}U_F,
\end{eqnarray}
where the starred functions are exactly the same as before.  Note that
the surface density $\Sigma$ must be axisymmetric in order that mass
be conserved around the orbit; this can be seen by vertically
integrating equation (\ref{rho}).  Therefore the $U$ quantities are
independent of $\phi$.  The disc thickness and the second moment are
non-axisymmetric, however, with
\begin{equation}
  \tilde{\sI}=\sI_*\,
  \left[f_1(\phi-\chi)\right]^{2/(6+x-2y)}
  \left[f_2(\phi-\chi)\right]^{-2(3-y)/(6+x-2y)}\Sigma U_H^2.
  \label{tildei}
\end{equation}
Here, as in Paper~I, the tilde denotes a non-axisymmetric quantity
prior to azimuthal averaging.  The azimuthal average is
$\sI=\langle\tilde{\sI}\rangle$.

Further propose that the velocities are all proportional to $\zeta$
and are of the form
\begin{eqnarray}
  v_r(\phi,\zeta)&=&f_3(\phi-\chi)\,r\Omega\zeta,\\
  v_\theta(\phi,\zeta)&=&f_4(\phi-\chi)\,r\Omega\zeta,\\
  v_\phi(\phi,\zeta)+rv_r(\phi,\zeta)\gamma^\prime\cos\beta&=&
  f_5(\phi-\chi)\,r\Omega\zeta,
\end{eqnarray}
as in Paper~I.

When this is substituted into equation (\ref{rho}), one obtains
\begin{equation}
  (3-y)f_2^\prime(\phi)f_1(\phi)-f_2(\phi)f_1^\prime(\phi)=
  (6+x-2y)f_2(\phi)f_1(\phi)f_4(\phi).
\end{equation}
Similarly, from equation (\ref{energy}) one obtains
\begin{eqnarray}
  f_2^\prime(\phi)&=&(\Gamma+1)f_2(\phi)f_4(\phi)
  +(\Gamma-1)f_2(\phi)
  \left(2\alpha\left[f_3(\phi)|\psi|\cos\phi\right]^2+
  2\alpha\left[f_4(\phi)+f_3(\phi)|\psi|\cos\phi\right]^2\right.\nonumber\\
  &&\left.+\alpha\left\{\left[f_4(\phi)+f_3(\phi)|\psi|\cos\phi\right]
  |\psi|\cos\phi-|\psi|\sin\phi-f_3(\phi)\right\}^2
  +\alpha\left[-q+f_5(\phi)|\psi|\cos\phi\right]^2+
  \alpha\left[f_5(\phi)\right]^2\right.\nonumber\\
  &&\left.+\left(\alpha_{\rm b}-{\textstyle{{2}\over{3}}}\alpha\right)
  \left[f_4(\phi)\right]^2-q^2\alpha f_1(\phi)\right).
\end{eqnarray}
Finally, from equations (\ref{vr})--(\ref{vphi}) one has
\begin{eqnarray}
  f_3^\prime(\phi)&=&f_4(\phi)f_3(\phi)+2f_5(\phi)+
  \left[1+(\alpha_{\rm b}+{\textstyle{{1}\over{3}}}\alpha)f_4(\phi)\right]
  f_2(\phi)|\psi|\cos\phi
  -\alpha f_2(\phi)f_3(\phi)(1+|\psi|^2\cos^2\phi)-
  \alpha f_2(\phi)|\psi|\sin\phi,\\
  f_4^\prime(\phi)&=&-f_3^\prime(\phi)|\psi|\cos\phi+
  2f_3(\phi)|\psi|\sin\phi+f_4(\phi)
  \left[f_4(\phi)+f_3(\phi)|\psi|\cos\phi\right]+1
  -\left[1+(\alpha_{\rm b}+{\textstyle{{1}\over{3}}}\alpha)
  f_4(\phi)\right]f_2(\phi)\nonumber\\
  &&-\alpha f_2(\phi)\left[f_4(\phi)+f_3(\phi)|\psi|\cos\phi\right]
  (1+|\psi|^2\cos^2\phi)
  +\alpha f_2(\phi)|\psi|^2\cos\phi\sin\phi,\\
  f_5^\prime(\phi)&=&f_4(\phi)f_5(\phi)-(2-q)f_3(\phi)-
  \alpha f_2(\phi)f_5(\phi)(1+|\psi|^2\cos^2\phi)+
  q\alpha f_2(\phi)|\psi|\cos\phi,
\end{eqnarray}
identical to equations (I.107)--(I.109).

As in Paper~I, define
\begin{equation}
  f_6(\phi-\chi)=\tilde{\sI}(\phi)/\sI,
\end{equation}
so that
\begin{equation}
  f_6^\prime(\phi)=-2f_4(\phi)f_6(\phi)
\end{equation}
and
\begin{equation}
  \langle f_6\rangle=1.
\end{equation}
Then the evolutionary equations for the warped disc have precisely the
same form as in Paper~I, viz.
\begin{equation}
  {{\partial\Sigma}\over{\partial t}}+
  {{1}\over{r}}{{\partial}\over{\partial r}}(rv\Sigma)=0
\end{equation}
and
\begin{equation}
  {{\partial}\over{\partial t}}(\Sigma r^2\Omega\,{\bmath\ell})+
  {{1}\over{r}}{{\partial}\over{\partial r}}
  (rv\Sigma r^2\Omega\,{\bmath\ell})=
  {{1}\over{r}}{{\partial}\over{\partial r}}
  \left(Q_1\sI r^2\Omega^2\,{\bmath\ell}+
  Q_2\sI r^3\Omega^2\,{{\partial{\bmath\ell}}\over{\partial r}}+
  Q_3\sI r^3\Omega^2\,{\bmath\ell}\times
  {{\partial{\bmath\ell}}\over{\partial r}}\right),
\end{equation}
where $v(r,t)$ is the mean radial velocity.  The dimensionless torque
coefficients $Q_1$ and $Q_4=Q_2+{\rm i}Q_3$ are again obtained from
\begin{equation}
  Q_1=\big\langle f_6\left[-q\alpha f_2-f_3f_5+
  \alpha f_2f_5|\psi|\cos\phi\right]\big\rangle
\end{equation}
and
\begin{equation}
  Q_4={{1}\over{|\psi|}}\big\langle{\rm e}^{{\rm i}\phi}f_6
  \left[f_3-{\rm i}f_3(f_4+f_3|\psi|\cos\phi)+
  {\rm i}\alpha f_2(f_4+f_3|\psi|\cos\phi)|\psi|\cos\phi-
  {\rm i}\alpha f_2f_3-{\rm i}\alpha f_2|\psi|\sin\phi\right]\big\rangle,
\end{equation}
in which $f_n$ stands for $f_n(\phi)$.

The relation between $\sI$ and $\Sigma$ is (from equation
\ref{tildei})
\begin{equation}
  \sI=Q_5\sI_*\,\Sigma U_H^2,
\end{equation}
where
\begin{equation}
  Q_5=\big\langle f_1^{2/(6+x-2y)}f_2^{-2(3-y)/(6+x-2y)}\big\rangle
\end{equation}
is a further coefficient to be evaluated numerically.  This relation
was missing in Paper~I because the thermal equilibrium of the disc had
not been considered.  It may be helpful to point out that this
relation, but with $Q_5=1$, corresponds to the usual relation between
the vertically integrated viscosity $\bar\nu\Sigma$ and the surface
density in a flat disc.  The two are linked in that case by
$\bar\nu\Sigma=\alpha\sI\Omega$.  One of the advantages of using $\sI$
is that it allows a consideration of inviscid discs (see Paper~I).
The coefficient $Q_5$ reflects the thickening of the disc in response
to increased viscous dissipation.

\section{Stability with respect to short-wavelength perturbations}

As in Paper~I, the evolutionary equations derived are of the general
form
\begin{equation}
  {{\partial\Sigma}\over{\partial t}}+
  {{1}\over{r}}{{\partial}\over{\partial r}}(rv\Sigma)=0,
\end{equation}
\begin{equation}
  {{\partial}\over{\partial t}}(\Sigma h\,{\bmath\ell})+
  {{1}\over{r}}{{\partial}\over{\partial r}}
  (rv\Sigma h\,{\bmath\ell})=
  {{1}\over{r}}{{\partial}\over{\partial r}}
  \left(g_1\,{\bmath\ell}+g_2\,r\,{{\partial{\bmath\ell}}\over{\partial r}}+
  g_3\,r\,{\bmath\ell}\times{{\partial{\bmath\ell}}\over{\partial r}}\right),
\end{equation}
where $v(r,t)$ is the mean radial velocity, $h(r)=r^2\Omega$ is the
specific angular momentum, and the torque components have the form
$g_i(r,\Sigma,|\psi|)$.  The system is of parabolic type and can be
thought of as a non-linear advection-diffusion-dispersion system.

In the case of a flat disc, the problem reduces to a non-linear
diffusion equation for the surface density.  It is well known
(Lightman \& Eardley 1974) that local thermal-viscous instability
occurs if $\partial(\bar\nu\Sigma)/\partial\Sigma<0$.  (It is assumed that
$h'>0$, because otherwise the orbital motion itself is unstable.)  In
the notation of the present paper, this criterion corresponds to
$\partial\sI/\partial\Sigma<0$.  It is important to find the
generalization of this condition to the case of a warped disc.

First eliminate $v$ to obtain
\begin{equation}
  h{{\partial}\over{\partial t}}(\Sigma\,{\bmath\ell})=
  {{1}\over{r}}{{\partial}\over{\partial r}}
  \left[g_1\,{\bmath\ell}+g_2r\,{{\partial{\bmath\ell}}\over{\partial r}}+
  g_3r\,{\bmath\ell}\times{{\partial{\bmath\ell}}\over{\partial r}}-
  \left({{\partial g_1}\over{\partial r}}-{{g_2|\psi|^2}\over{r}}\right)
  {{h}\over{h'}}\,{\bmath\ell}\right].
\end{equation}
Consider the stability of any given solution of this equation with
respect to linear Eulerian perturbations
$(\delta\Sigma,\delta{\bmath\ell})$.  Note that
\begin{equation}
  \delta g_i={{\partial g_i}\over{\partial\Sigma}}\,\delta\Sigma+
  {{\partial g_i}\over{\partial|\psi|}}\,\delta|\psi|,
\end{equation}
while
\begin{equation}
  \delta|\psi|={{r^2}\over{|\psi|}}
  {{\partial{\bmath\ell}}\over{\partial r}}\cdot
  {{\partial{\delta\bmath\ell}}\over{\partial r}}
\end{equation}
and ${\bmath\ell}\cdot\delta{\bmath\ell}=0$.  The perturbed angular
momentum equation has the form
\begin{eqnarray}
  h{{\partial}\over{\partial t}}
  (\delta\Sigma\,{\bmath\ell}+\Sigma\,\delta{\bmath\ell})&=&
  {{1}\over{r}}{{\partial}\over{\partial r}}
  \left[\delta g_1\,{\bmath\ell}+g_1\,\delta{\bmath\ell}+
  \delta g_2r\,{{\partial{\bmath\ell}}\over{\partial r}}+
  g_2r\,{{\partial\delta{\bmath\ell}}\over{\partial r}}+
  \delta g_3r\,{\bmath\ell}\times{{\partial{\bmath\ell}}\over{\partial r}}  
  +g_3r\,\delta{\bmath\ell}\times{{\partial{\bmath\ell}}\over{\partial r}}+
  g_3r\,{\bmath\ell}\times{{\partial\delta{\bmath\ell}}\over{\partial r}}
  \right.\nonumber\\
  &&\left.\qquad\qquad-\left({{\partial\delta g_1}\over{\partial r}}-
  {{\delta(g_2|\psi|^2)}\over{r}}\right){{h}\over{h'}}\,{\bmath\ell}-
  \left({{\partial g_1}\over{\partial r}}-{{g_2|\psi|^2}\over{r}}\right)
  {{h}\over{h'}}\,\delta{\bmath\ell}\right].
  \label{perturbed}
\end{eqnarray}
Now suppose the perturbations vary much more rapidly with radius and
time than the unperturbed solution, so that they have the form
\begin{equation}
  \exp\left(-{\rm i}\int\omega\,{\rm d}t+{\rm i}\int k\,{\rm d}r\right)
\end{equation}
multiplied by more slowly varying functions.  Here $|\omega
t|,|kr|\gg1$ and one may formally let $k\to\infty$ (although in
reality the equations become invalid when $k$ is comparable to $1/H$).
If one adopts the arbitrary scaling $|\delta{\bmath\ell}|=O(k^{-1})$,
then $\delta\Sigma=O(1)$, $\delta|\psi|=O(1)$ and $\delta g_i=O(1)$,
while $\omega=O(k^2)$.

Equation (\ref{perturbed}) should then be projected on to the
orthonormal basis $({\bmath\ell},{\bmath m},{\bmath n})$ defined by
$r(\partial{\bmath\ell}/\partial r)=|\psi|\,{\bmath m}$ and
${\bmath\ell}\times{\bmath m}={\bmath n}$.  Writing
$\delta{\bmath\ell}=\delta m\,{\bmath m}+\delta n\,{\bmath n}$, one
obtains, at leading order in $k$,
\begin{eqnarray}
  -{\rm i}\omega hr\,\delta\Sigma&=&k^2{{h}\over{h'}}\,\delta g_1,\\
  -{\rm i}\omega hr\Sigma\,\delta m&=&{\rm i}k|\psi|\,\delta g_2-
  k^2g_2r\,\delta m+k^2g_3r\,\delta n-
  {\rm i}k|\psi|{{h}\over{rh'}}\,\delta g_1,\\
  -{\rm i}\omega hr\Sigma\,\delta n&=&-k^2g_2r\,\delta n+
  {\rm i}k|\psi|\,\delta g_3-k^2g_3r\,\delta m,
\end{eqnarray}
while $\delta|\psi|={\rm i}kr\,\delta m$.  The solvability condition
yields a dispersion relation in the form of a cubic equation for
$\omega/k^2$; instability occurs if any of the roots has a positive
imaginary part, since then short-wavelength disturbances grow
exponentially in time.

It has been shown that, under the assumptions of this paper, the
torque coefficients have the form
\begin{equation}
  g_i=Q_i(|\psi|)\sI r^2\Omega^2,
\end{equation}
with
\begin{equation}
  \sI=Q_5(|\psi|)\Sigma^pF(r),
\end{equation}
where the exponent $p=5/3$ (Thomson) or $10/7$ (Kramers), and $F(r)$
is some function.  Define the dimensionless growth rate $s$ by
\begin{equation}
  -{\rm i}\omega=s{{\sI}\over{\Sigma}}\Omega k^2;
\end{equation}
then the cubic dispersion relation corresponds to
\begin{equation}
  \det\left[\matrix{s-apQ_1&-a(Q_1Q_5)'/Q_5&0\cr
  (Q_2-aQ_1)p|\psi|&s+Q_2+(Q_2Q_5)'|\psi|/Q_5-a(Q_1Q_5)'|\psi|/Q_5&-Q_3\cr
  pQ_3|\psi|&Q_3+(Q_3Q_5)'|\psi|/Q_5&s+Q_2\cr}\right]=0,
\end{equation}
where $a=h/(rh')$, and a prime on a $Q$ coefficient denotes
differentiation with respect to $|\psi|$.

If the disc is initially flat ($|\psi|=0$), the dispersion relation
has roots
\begin{equation}
  s=apQ_1,\qquad s=-Q_2\pm{\rm i}Q_3.
\end{equation}
For stability, one requires $apQ_1<0$ and $Q_2>0$.  Note that $p>0$ (or,
rather, $ap>0$) is the usual criterion for thermal-viscous stability
as discussed above.  For a flat disc, the $Q$ coefficients always
satisfy $Q_1<0$ and $Q_2>0$, except possibly for unphysical rotation
laws.  The condition $Q_2>0$ means that the effective diffusion
coefficient for a small warp introduced into the disc is positive.

For an initially warped disc, the dispersion relation contains
$|\psi|$ as a parameter and its roots must be determined numerically
along with the $Q$ coefficients.  Instability occurs when a root
exists with ${\rm Re}(s)>0$.

\section{Numerical results}

\begin{figure}
\centerline{\epsfbox{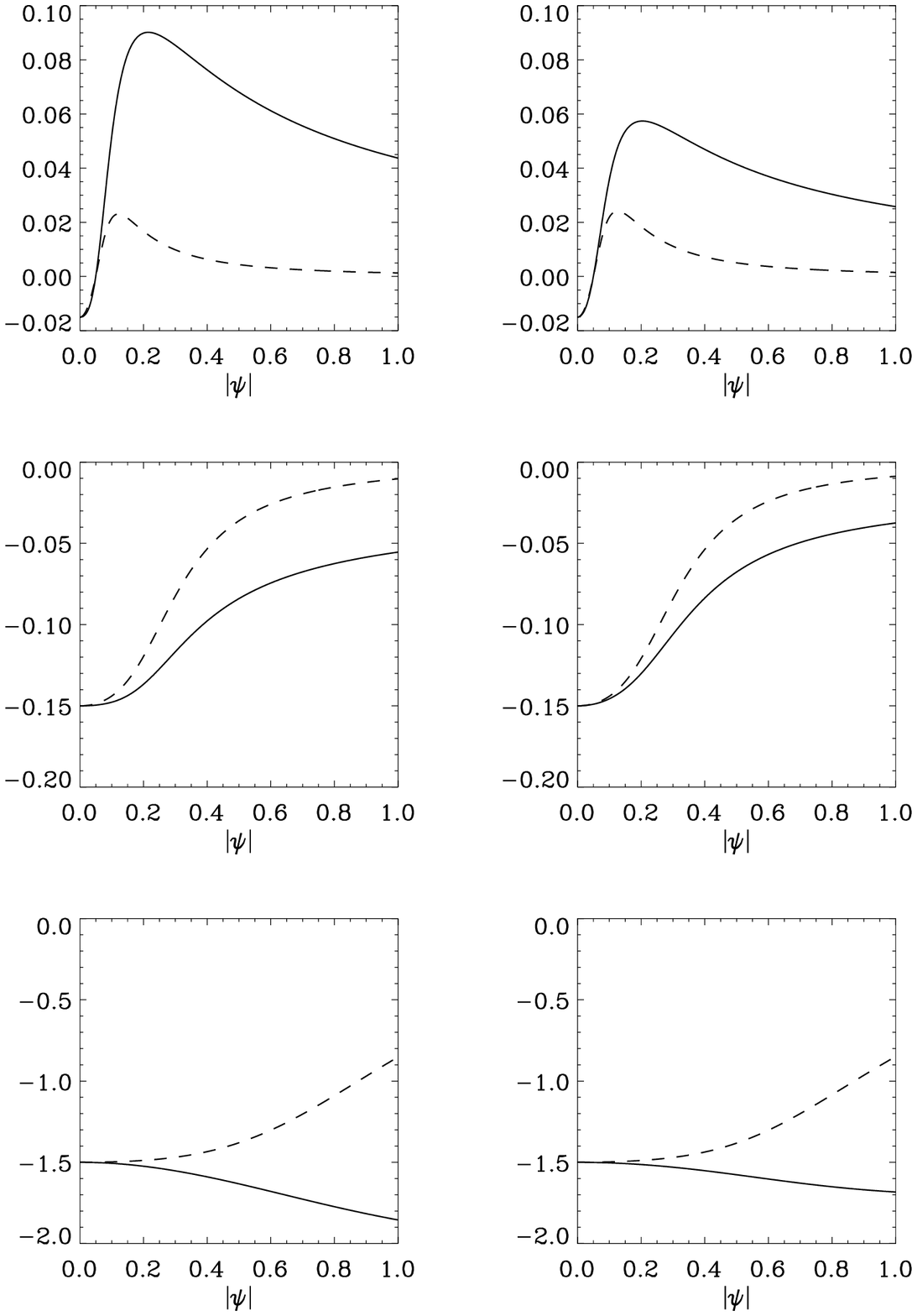}}
\caption{The variation of the coefficient $Q_1$ (dashed lines) and the
  product $Q_1Q_5$ (solid lines) with the amplitude $|\psi|$ of the
  warp, for a Keplerian disc with $\Gamma=5/3$ and no bulk viscosity.
  The left-hand panels are for Thomson opacity, the right-hand panels
  for Kramers opacity.  The upper, middle and lower panels correspond
  to $\alpha=0.01$, $0.1$ and $1$ respectively.}
\end{figure}

\begin{figure}
\centerline{\epsfbox{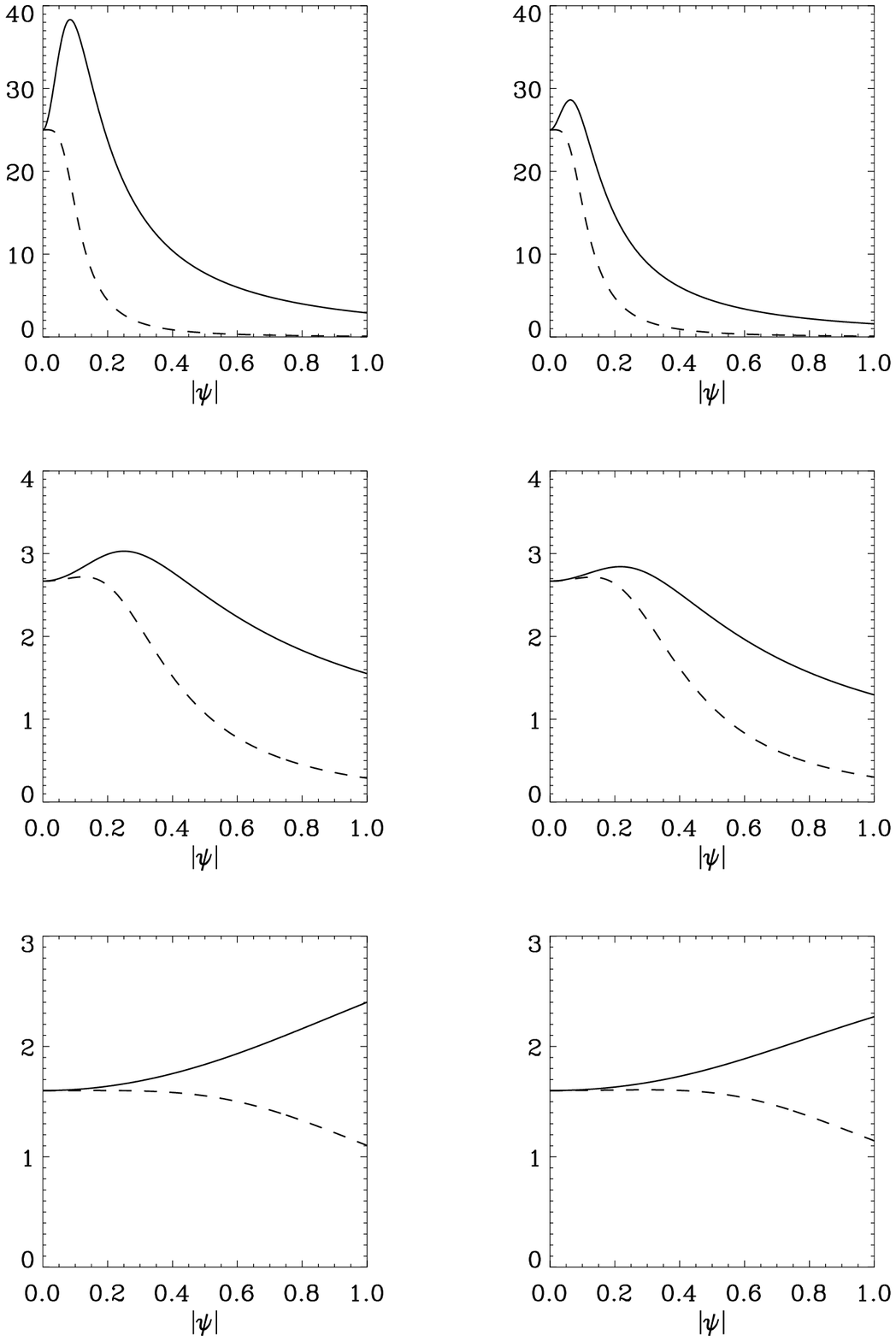}}
\caption{The coefficient $Q_2$ (dashed lines) and the product $Q_2Q_5$
  (solid lines), as for Fig.~1.}
\end{figure}

\begin{figure}
\centerline{\epsfbox{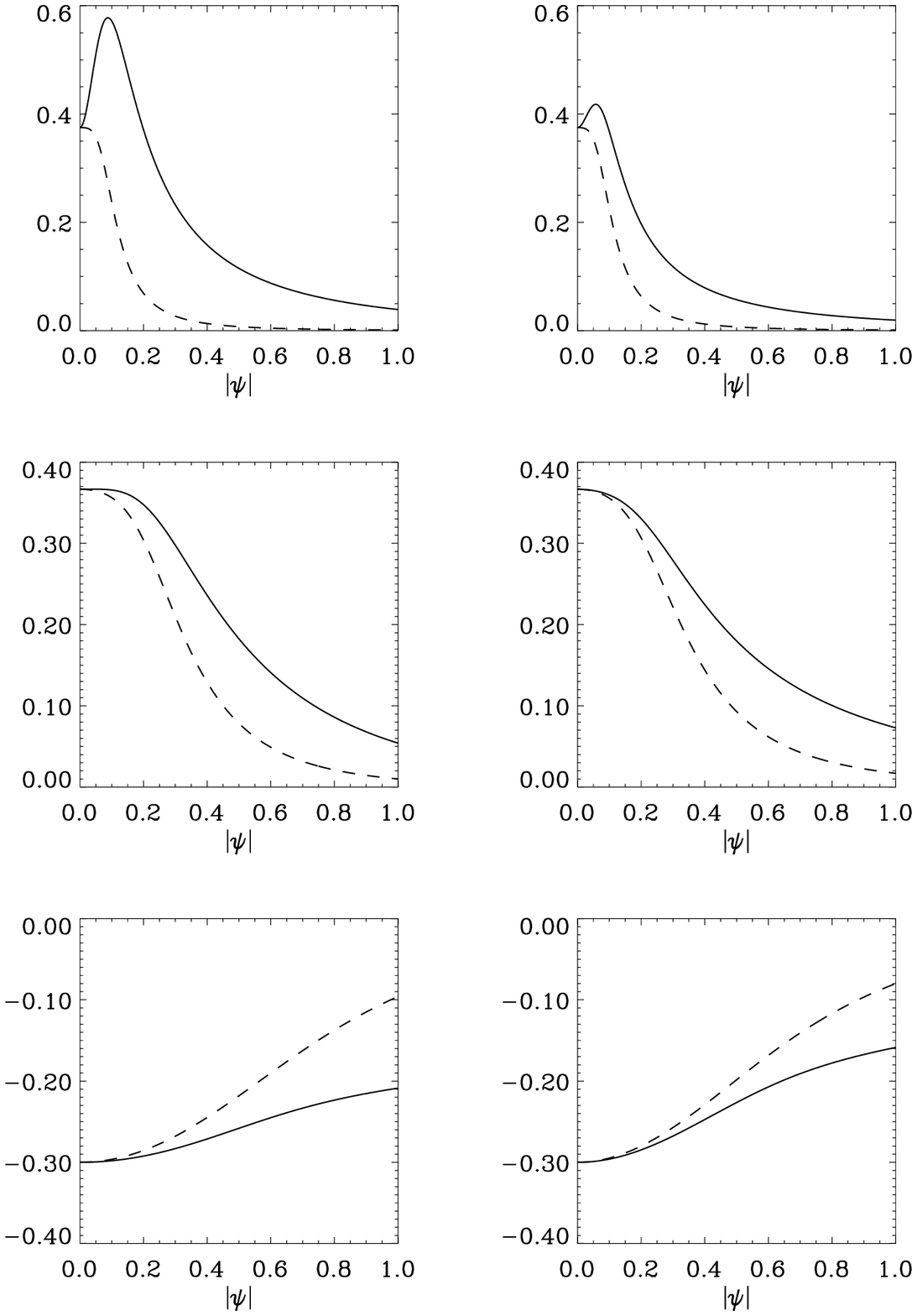}}
\caption{The coefficient $Q_3$ (dashed lines) and the product $Q_3Q_5$
  (solid lines), as for Fig.~1.}
\end{figure}

The numerical evaluation of the dimensionless torque coefficients
proceeds similarly to Paper~I.  Attention is restricted here to the
case of a Keplerian disc ($q=3/2$) with adiabatic exponent
$\Gamma=5/3$ and vanishing bulk viscosity ($\alpha_{\rm b}=0$).

In Figs~1--3 the coefficients $Q_1$, $Q_2$ and $Q_3$, and also the
products $Q_1Q_5$, $Q_2Q_5$ and $Q_3Q_5$, are plotted as functions of
the amplitude $|\psi|$ of the warp.  Three values for the viscosity
coefficient, $\alpha=0.01$, $0.1$ and $1$, are investigated, and both
Thomson and Kramers opacity are considered.  The coefficients $Q_1$,
$Q_2$ and $Q_3$ depend very little on the opacity law and are very
similar to their values for the polytropic disc considered in Paper~I.
However, the coefficient $Q_5$ increases rapidly with increasing
amplitude, reflecting the thickening of the disc in response to
increased viscous dissipation.  The precise behaviour of $Q_5$ depends
to some degree on the opacity law.  The products $Q_1Q_5$, $Q_2Q_5$
and $Q_3Q_5$ reflect the true variation of the three torque components
with the amplitude of the warp.

If $\alpha$ is large, say $\alpha=1$, the variation of $Q_1Q_5$,
$Q_2Q_5$ and $Q_3Q_5$ with amplitude is modest.  However, for smaller
$\alpha$ the variation can be significant.  In particular, if
$\alpha\la0.05$, $Q_1$ can eventually become positive, meaning that
the usual viscous torque parallel to ${\bmath\ell}$ is reversed.
However, as noted in Paper~I, the disc becomes highly non-axisymmetric
in this limit and the validity of the solution appears questionable.

The roots of the local dispersion relation for short-wavelength
perturbations can be calculated from this information.  The system is
found to be stable for $0\le|\psi|\le1$ when $\alpha=0.1$ or
$\alpha=1$.  However, for smaller $\alpha$, instability is found for
sufficiently large $|\psi|$.  The critical values of $|\psi|$ for
various values of $\alpha$ are listed in Table~1.  Instability
usually, although not invariably, occurs where $Q_1$ is already
positive.

\begin{table}
  \caption{Critical warp amplitude for local instability in a Keplerian disc.}
  \begin{tabular}{lcc}
    $\alpha$&$|\psi|$ (Thomson)&$|\psi|$ (Kramers)\\
    &&\\
    0.01&0.19&0.17\\
    0.02&0.29&0.24\\
    0.03&0.37&0.31\\
    0.04&0.47&0.39\\
    0.05&0.63&0.47\\
  \end{tabular}
\end{table}

\section{Summary and discussion}

\subsection{Summary of results}

Under the assumptions of this paper, the non-linear dynamics of a
warped accretion disc may be described by one-dimensional conservation
equations for mass,
\begin{equation}
  {{\partial\Sigma}\over{\partial t}}+
  {{1}\over{r}}{{\partial}\over{\partial r}}(rv\Sigma)=0,
\end{equation}
and angular momentum,
\begin{equation}
  {{\partial}\over{\partial t}}(\Sigma r^2\Omega\,{\bmath\ell})+
  {{1}\over{r}}{{\partial}\over{\partial r}}
  (rv\Sigma r^2\Omega\,{\bmath\ell})=
  {{1}\over{r}}{{\partial{\bmath G}}\over{\partial r}}+{\bmath T}.
\end{equation}
Here $\Sigma(r,t)$ is the surface density, $v(r,t)$ the mean radial
velocity, $\Omega(r)$ the orbital angular velocity and
${\bmath\ell}(r,t)$ the tilt vector.  The internal torque is
$2\pi{\bmath G}(r,t)$, given by
\begin{equation}
  {\bmath G}=Q_1\sI r^2\Omega^2\,{\bmath\ell}+
  Q_2\sI r^3\Omega^2\,{{\partial{\bmath\ell}}\over{\partial r}}+
  Q_3\sI r^3\Omega^2\,{\bmath\ell}\times
  {{\partial{\bmath\ell}}\over{\partial r}},
\end{equation}
where $\sI(r,t)$ is the azimuthally averaged second vertical moment of
the density.  Finally, ${\bmath T}(r,t)$ represents any external
torque due to self-gravitation, radiation forces, tidal forcing, etc.
The dimensionless coefficients $Q_1$, $Q_2$ and $Q_3$ depend on the
rotation law and the shear viscosity, and also, in the non-linear
theory, on the adiabatic exponent, the opacity law, the bulk viscosity
and the amplitude $|\psi|=r|\partial{\bmath\ell}/\partial r|$ of the
warp.

For an ideal gas and a power-law opacity function, the relation
between $\sI$ and $\Sigma$ is
\begin{equation}
  \sI=Q_5\sI_*\Sigma^{(10+3x-2y)/(6+x-2y)}
  (q^2\alpha)^{2/(6+x-2y)}\Omega^{-2(5-2y)/(6+x-2y)}
  \left({{\mu_{\rm m}m_{\rm H}}\over{k}}\right)^{-2(4-y)/(6+x-2y)}
  \left({{16\sigma}\over{3C_\kappa}}\right)^{-2/(6+x-2y)},
\end{equation}
where $Q_5$ is a further dimensionless coefficient, equal to unity for
a flat disc.  Rosseland mean opacities for a standard `solar' chemical
composition have been computed by, e.g., Iglesias \& Rogers (1996),
who tabulate $\log\kappa_{\rm R}$ as a function of $\log T$ and $\log
R=\log\rho-3\log T+18$ (all quantities expressed in CGS units).
Power-law opacity functions of the Thomson and Kramers type offer a
fair approximation to two large sectors of this table.  Fiducial
values of $C_\kappa=0.33$ for the Thomson region (based on $\log T=7$
and $\log R=-7$) and $C_\kappa=4.5\times10^{24}$ for the Kramers
region (based on $\log T=4.5$ and $\log R=-3$) will be adopted.  The
errors in these quantities, especially the latter, are quite large,
but fortunately the dependence of $\sI$ on $C_\kappa$ is very weak
($\sI\propto C_\kappa^{1/3}$ for Thomson and $\sI\propto C_\kappa^{1/7}$
for Kramers), so $C_\kappa$ need not be defined very accurately.

For a Keplerian disc ($q=3/2$) with $\mu_{\rm m}=0.6$,
this gives
\begin{equation}
  \sI\approx6.5\times10^{11}\,Q_5\,\alpha^{1/3}\Sigma^{5/3}\Omega^{-5/3}
\end{equation}
for Thomson opacity and
\begin{equation}
  \sI\approx4.4\times10^{12}\,Q_5\,\alpha^{1/7}\Sigma^{10/7}\Omega^{-12/7}
\end{equation}
for Kramers opacity, where all quantities are referred to CGS units.
The numerical coefficients quoted here involve estimated uncertainties
of 10--20\% owing to variations in $C_\kappa$ and $\mu_{\rm m}$,
depending on how the power-law opacity regions are delineated.  Minor
adjustments may also be required to account for the effects of finite
optical thickness and/or external irradiation, as discussed in the
Appendix.

The $Q$ coefficients can readily be calculated numerically by solving
a set of ODEs in azimuth.  These are
\begin{eqnarray}
  f_1^\prime&=&(3-y)f_2^\prime f_1/f_2-(6+x-2y)f_1f_4,\\
  f_2^\prime&=&(\Gamma+1)f_2f_4+(\Gamma-1)f_2
  \left\{2\alpha\left(f_3|\psi|\cos\phi\right)^2+
  2\alpha\left(f_4+f_3|\psi|\cos\phi\right)^2\right.\nonumber\\
  &&\left.+\alpha\left[\left(f_4+f_3|\psi|\cos\phi\right)
  |\psi|\cos\phi-|\psi|\sin\phi-f_3\right]^2+
  \alpha\left(-q+f_5|\psi|\cos\phi\right)^2
  +\alpha f_5^2+
  \left(\alpha_{\rm b}-{\textstyle{{2}\over{3}}}\alpha\right)f_4^2-
  q^2\alpha f_1\right\},\\
  f_3^\prime&=&f_4f_3+2f_5+
  \left[1+(\alpha_{\rm b}+{\textstyle{{1}\over{3}}}\alpha)f_4\right]
  f_2|\psi|\cos\phi-\alpha f_2f_3(1+|\psi|^2\cos^2\phi)
  -\alpha f_2|\psi|\sin\phi,\\
  f_4^\prime&=&-f_3^\prime|\psi|\cos\phi+2f_3|\psi|\sin\phi+
  f_4\left(f_4+f_3|\psi|\cos\phi\right)+1-
  \left[1+(\alpha_{\rm b}+{\textstyle{{1}\over{3}}}\alpha)f_4\right]
  f_2\nonumber\\
  &&-\alpha f_2\left(f_4+f_3|\psi|\cos\phi\right)
  (1+|\psi|^2\cos^2\phi)+\alpha f_2|\psi|^2\cos\phi\sin\phi,\\
  f_5^\prime&=&f_4f_5-(2-q)f_3-\alpha f_2f_5(1+|\psi|^2\cos^2\phi)+
  q\alpha f_2|\psi|\cos\phi,\\
  f_6^\prime&=&-2f_4f_6,
\end{eqnarray}
in which $f_n$ stands for $f_n(\phi)$.  These are subject to periodic
boundary conditions $f_n(2\pi)=f_n(0)$ and the normalization condition
$\langle f_6\rangle=1$, where the angle brackets denote azimuthal
averaging.  The required coefficients follow from
\begin{eqnarray}
  Q_1&=&\big\langle f_6\left[-q\alpha f_2-f_3f_5+
  \alpha f_2f_5|\psi|\cos\phi\right]\big\rangle,\\
  Q_2+{\rm i}Q_3&=&{{1}\over{|\psi|}}\big\langle{\rm e}^{{\rm i}\phi}
  f_6\left[f_3-{\rm i}f_3(f_4+f_3|\psi|\cos\phi)+
  {\rm i}\alpha f_2(f_4+f_3|\psi|\cos\phi)|\psi|\cos\phi
  -{\rm i}\alpha f_2f_3-{\rm i}\alpha f_2|\psi|\sin\phi\right]\big\rangle,\\
  Q_5&=&\big\langle f_1^{2/(6+x-2y)}f_2^{-2(3-y)/(6+x-2y)}\big\rangle.
\end{eqnarray}

Apart from the torque component with coefficient $Q_3$, a direct
correspondence with the notation of Pringle (1992) can be made by
identifying
\begin{eqnarray}
  Q_1\sI\Omega&\leftrightarrow&\nu_1\Sigma
  {{{\rm d}\ln\Omega}\over{{\rm d}\ln r}},\\
  Q_2\sI\Omega&\leftrightarrow&\nu_2\Sigma.
\end{eqnarray}
Therefore $\nu_1\propto-Q_1Q_5$ and $\nu_2\propto Q_2Q_5$.  The
numerical evaluation of the $Q$ coefficients (Section~4) demonstrates
that the qualitative behaviour of $\nu_1$ and $\nu_2$ depends on the
value of the small-scale viscosity coefficient $\alpha$.  For large
$\alpha$, say $\alpha=1$, both $\nu_1$ and $\nu_2$ increase with
increasing amplitude.  For smaller $\alpha$, $\nu_1$ decreases with
increasing amplitude and may even become negative; while $\nu_2$ first
increases and then decreases.  Also important is the variation of
$\nu_1$ and $\nu_2$ with $\Sigma$, which is the same as for a flat
disc.

\subsection{Discussion}

The non-linear theory derived in this paper is suitable for many
applications.  Some comment is warranted on the meaning of the
solution and the conditions under which it is expected to be valid.

Starting from the basic fluid-dynamical equations in three dimensions,
I have derived one-dimensional evolutionary equations using a
consistent asymptotic expansion that is based on the thinness of the
disc, $H/r=O(\epsilon)$, and the consequent separation of the orbital
and evolutionary time-scales.  The theory is fully non-linear in the
sense that warp amplitudes $|\psi|$ of order unity are allowed for.
Although this `solution' is not an exact solution of the Navier-Stokes
and accompanying equations, the theory presented here is
asymptotically exact, meaning that it differs from an exact solution
by an amount that tends to zero as $O(\epsilon)$ in the limit
$\epsilon\to0$.  It may be noted that any attempt to define the tilt
vector of a disc of non-zero thickness to an accuracy better than
$O(\epsilon)$ is somewhat arbitrary.

The validity of the solution should be questioned when this error,
although formally $O(\epsilon)$, is numerically large and accumulates
rapidly.  This is expected to occur for a Keplerian disc when
$\alpha\la H/r$, when a resonance occurs, as described in Paper~I.  In
this case $\epsilon$ is not small enough for the asymptotic analysis
to apply.  On the other hand, the SPH simulations of Nelson \&
Papaloizou (1999) suggest that, even when $\alpha\la H/r$, the
evolution of warps again becomes diffusive in character when the
amplitude is large.  This is consistent with the observation made in
Paper~I that the resonant behaviour is weakened at large amplitude; it
can also be seen from the upper panels of Fig.~2, where the resonant
peak in the coefficient $Q_2$ is located at small amplitude only.  It
would be useful to make more detailed comparisons between SPH
simulations and the present theory.

Although Nelson \& Papaloizou (1999) found shock-like features in some
of their simulations, this occurred when the wavelength of the bending
disturbance was comparable to the disc thickness so that a collision
occurred between oppositely directed horizontal flows.  The present
theory does not allow for such short-wavelength disturbances.  In any
case, it is not expected that strict discontinuities will occur in a
solution of the Navier-Stokes equation.

A further issue is the stability of the solution.  As shown in
Section~3, if a flat disc is stable to the thermal-viscous
instability, it usually remains viscously stable when it is
warped.\footnote{The two specific opacity laws considered in this
  paper allow for flat discs that are stable.  The possibility of
  thermal-viscous instability due to incomplete hydrogen ionization
  was not considered.} An exception arises if $\alpha$ is small, say
$\alpha\la0.05$, when the disc may become locally unstable at
sufficiently large warp amplitude.  It has been noted, however, that
the deformation of the disc is more extreme in this limit, and the
induced motions are more likely to be hydrodynamically unstable.
Therefore the viscous instability, whose physical origin is unclear,
may not occur in practice.  It is not evident what the true outcome of
a large-amplitude warp in a low-viscosity disc would be, but the
parametric instability discussed by Gammie, Goodman \& Ogilvie (2000)
is likely to be important.

Provided that $\alpha$ is not so small as this, the principal
remaining uncertainty in this analysis concerns the modelling of the
turbulent stress tensor.  The simplest assumption, adopted here as in
almost all theoretical work on accretion discs (and, implicitly, in
SPH simulations), is that the stress behaves similarly to an isotropic
viscosity.  Measurements of the damping rate, caused by
magnetohydrodynamic turbulence, of shearing motions of the type
induced in a warped disc (Torkelsson et~al. 2000) broadly support this
hypothesis, but more numerical and theoretical effort is required in
this area.

The very innermost part of a disc around a neutron star or black hole
is dominated by radiation pressure and Thomson scattering opacity.
This case would require a treatment of the radiation-hydrodynamic
equations and is beyond the scope of this paper.  However, one can
anticipate that the evolutionary equations are identical in form and
that the appropriate $\sI$--$\Sigma$ relation for a flat disc is
modified with a $Q_5$ coefficient in a similar way.  However, the
detailed variation of the $Q$ coefficients would require a further
investigation.

Discs in binary systems are subject to tidal forces that cause a
distortion of the flow, possibly including spiral shocks.  The
interaction of such distortions with a warp requires further
investigation.  If spiral shocks are the dominant mode of angular
momentum transport in the disc, the `viscous' theory developed here
and elsewhere may not be relevant.

In conclusion, this paper provides the logical generalization of the
alpha theory of Shakura \& Sunyaev (1973) to the case of a
time-dependent warped accretion disc.  It is derived directly from the
basic fluid-dynamical equations, and is valid even for warps of large
amplitude.  The scheme is suitable for practical applications: it is
expressed in physical units, it requires the solution of only
one-dimensional equations from which the fast orbital time-scale has
been eliminated, and it allows numerous external effects to be
included in terms of the torque that they exert on the disc.  Such
applications will be presented in future publications.

\section*{Acknowledgments}

I thank Jim Pringle and Henk Spruit for helpful discussions.  I
acknowledge the support of Clare College, Cambridge through a research
fellowship, and of the European Commission through the TMR network
`Accretion on to Black Holes, Compact Stars and Protostars' (contract
number ERBFMRX-CT98-0195).  The Appendix on the effects of irradiation
was added at the suggestion of an anonymous referee.

\appendix

\section{Effects of finite optical thickness and external irradiation}

In Sections~2.3 and~2.4 the vertical structure of the disc was
determined by applying the `zero boundary conditions' $\rho=T=0$ at
the surface.  This is correct in the limit of a highly optically thick
disc.  In practice the optical thickness is usually large but
certainly finite, and this leads to a small correction to the
dynamical equations through a change in the dimensionless quantity
$\sI_*$.  More importantly, the disc may be subject to an external
irradiating flux, which can affect the vertical structure.  This is
particularly important for discs around neutron stars and black holes.
If the disc is warped it is likely to intercept a significant fraction
of the radiation emitted by the central X-ray source.

The effects of finite optical thickness and external irradiation may
be estimated from a simple model, derived in part from a more detailed
analysis by Dubus et~al. (1999).  It is supposed that the equations of
Section~2.2 extend up to a photospheric surface $\zeta=\zeta_{\rm s}$,
above which $F=F_{\rm s}={\rm constant}$ and $T=T_{\rm s}={\rm
  constant}$.  The constancy of $F$ implies that there is negligible
dissipation of energy in the atmosphere, while the constancy of $T$ is
a good approximation to more detailed atmospheric models.  The
possibility of a hot corona is neglected here.

The photospheric boundary conditions are taken to be
\begin{equation}
  F_{\rm s}+\sigma T_{\rm irr}^4=\sigma T_{\rm s}^4,
\end{equation}
and
\begin{equation}
  \tau_{\rm s}={{2}\over{3}},
\end{equation}
where $T_{\rm irr}$ is the irradiation temperature and
\begin{equation}
  \tau_{\rm s}=\int_{\zeta_{\rm s}}^\infty\kappa_{\rm R}\rho
  \,r\,{\rm d}\zeta
\end{equation}
the optical depth of the photosphere.  To a good approximation this
equates to
\begin{equation}
  {{2}\over{3}}=C_\kappa\rho_{\rm s}^{1+x}T_{\rm s}^yr
  \int_0^\infty\exp\left[-(1+x)
  \left({{\zeta-\zeta_{\rm s}}\over{h_{\rm s}}}\right)\right]\,
  {\rm d}(\zeta-\zeta_{\rm s}),
\end{equation}
where $h_{\rm s}$ is the density scale-height (in terms of the
variable $\zeta$) at the photosphere, given by
\begin{equation}
  {{kT_{\rm s}}\over{\mu_{\rm m}m_{\rm H}}}{{1}\over{h_{\rm s}}}=
  \left(-{{1}\over{\rho}}{{\partial p}\over{\partial\zeta}}\right)
  _{\rm s}.
\end{equation}
This may be rearranged into the form
\begin{equation}
  \rho_{\rm s}^{1+x}T_{\rm s}^y=\left({{1+x}\over{8\sigma}}\right)
  \left({{\mu_{\rm m}m_{\rm H}}\over{k}}\right)
  \left({{16\sigma}\over{3C_\kappa}}\right)
  \left(-{{1}\over{\rho r}}{{\partial p}\over{\partial\zeta}}\right)
  _{\rm s}.
\end{equation}

For a flat disc, the analysis of Section~2.3 proceeds as before, but
now with boundary conditions
\begin{equation}
  \delta F_*(\zeta_{{\rm s}*})+T_{{\rm irr}*}^4=
  [T_*(\zeta_{{\rm s}*})]^4
\end{equation}
and
\begin{equation}
  [\rho_*(\zeta_{{\rm s}*})]^{1+x}[T_*(\zeta_{{\rm s}*})]^{1+y}=
  {{1}\over{8}}\delta(1+x)\zeta_{{\rm s}*},
\end{equation}
where
\begin{equation}
  T_{\rm irr}=T_{{\rm irr}*}\,U_T,
\end{equation}
and
\begin{equation}
  \delta={{U_F}\over{\sigma U_T^4}}
\end{equation}
is a small dimensionless parameter.  
  When the solution is obtained,
the optical depth at the mid-plane can be determined from
\begin{equation}
  \tau_{\rm c}={{2}\over{3}}+
  \int_0^{\zeta_{\rm s}}\kappa_{\rm R}\rho\,r\,{\rm d}\zeta=
  {{2}\over{3}}+{{16}\over{3\delta}}\int_0^{\zeta_{{\rm s}*}}
  \rho_*^{1+x}T_*^y\,{\rm d}\zeta_*.
\end{equation}
Therefore $\delta$ is an approximate inverse measure of $\tau_{\rm
  c}$.

The computed values of $\sI_*$ for some representative values of
$\delta$ and $T_{{\rm irr}*}$ are given in Table~A1.  Also given are
the optical depth at the mid-plane $\tau_{\rm c}$, the surface
temperature $T_{{\rm s}*}$ and the central temperature $T_{{\rm c}*}$.
The contribution of the atmosphere to $\Sigma$ and $\sI$ has been
neglected.  It can be seen that a noticeable change in $\sI_*$ occurs,
as expected, when the optical thickness is not large, or when the
irradiation temperature exceeds the central temperature of the
non-irradiated solution.  However, the fractional change is less than
$30\%$ even for a strongly irradiated disc with $\tau_{\rm c}\sim10$.

\begin{table}
  \caption{Effects of finite optical thickness and external irradiation on the second moment in the case of a flat disc with Kramers opacity.}
  \begin{tabular}{lccccc}
    $\delta$&$T_{{\rm irr}*}$&$T_{{\rm s}*}$&$T_{{\rm c}*}$&
    $\tau_{\rm c}$&$\sI_*$\\
    &&\\
    0.001&0&0.137&0.802&2740&0.708\\
    0.01&0&0.244&0.803&274&0.699\\
    0.1&0&0.437&0.814&27.0&0.622\\
    0.001&0.3&0.303&0.803&2710&0.704\\
    0.01&0.3&0.329&0.804&271&0.691\\
    0.1&0.3&0.460&0.816&26.8&0.614\\
    0.001&1.0&1.000&1.027&723&0.899\\
    0.01&1.0&1.001&1.031&77.7&0.753\\
    0.1&1.0&1.013&1.049&9.61&0.503\\
  \end{tabular}
\end{table}

It has not been found possible to give an exact solution for the
vertical structure of a {\it warped} disc including the photospheric
boundary conditions.  The difficulties are especially pronounced in
the case of a disc irradiated by the central X-ray source, since the
irradiation then varies strongly with azimuth (especially if the disc
is partially self-shadowed) and is one-sided, breaking the symmetry of
the vertical structure.  However, for $\alpha<1$ the thermal
time-scale is longer than the orbital time-scale and it is appropriate
to average the irradiation over the orbital motion of the fluid.  This
restores the symmetry about the mid-plane and suggests that the effect
on the $\sI$--$\Sigma$ relation will be similar to that found above
for a flat disc.

\label{lastpage}

\end{document}